# Secure Watermarking Scheme for Color Image Using Intensity of Pixel and LSB Substitution

Nagaraj V. Dharwadkar and B. B. Amberker

**Abstract**— In this paper a novel spatial domain LSB based watermarking scheme for color Images is proposed. The proposed scheme is of type blind and invisible watermarking. Our scheme introduces the concept of storing variable number of bits in each pixel based on the actual color value of pixel. Equal or higher the color value of channels with respect to intensity of pixel stores higher number of watermark bits. The Red, Green and Blue channel of the color image has been used for watermark embedding. The watermark is embedded into selected channels of pixel. The proposed method supports high watermark embedding capacity, which is equivalent to the size of cover image. The security of watermark is preserved by permuting the watermark bits using secret key. The proposed scheme is found robust to various image processing operations such as image compression, blurring, salt and pepper noise, filtering and cropping.

—————————— ◆ ——————————

## 1 INTRODUCTION

DIGITIZATION and new coding formats, together with the development of the Web, have led to a world where electronic distribution of image, video and audio to end users has become a reality. This reality however has at the same time led to increased concern about the protection of the rights of owners of the content that is distributed in electronic form. Digital right management (DRM) is the term for commercial, legal, and technical measures that can be taken to ensure that rights of owners of digital content are respected [1]. One approach to address the problems of DRM is watermarking technique. The Digital watermarking has been proposed as a way to identify the source of creator, and distributor of image [2].

The process of watermarking involves the modification of original information data to embed watermark information. Various watermarking techniques have been developed. However, these techniques can be grouped into two classes: spatial domain and frequency domain. The spatial domain methods are to embed the watermark by directly modifying the pixel values of the original image. LSB embedding is one of algorithm that uses spatial domain. When LSB is applied in the spatial or temporal domains, these approaches modify the Least Significant Bits (LSB) of the host data. The invisibility of the watermark is achieved on the assumption that the LSB data are visually insignificant [3]. The watermark is generally recovered using knowledge of the PN sequence (and perhaps other secret keys, like watermark location) and the statistical properties of the embedding process. In the literature there are many LSB techniques. In [4], the watermark is inserted in to the LSB of image pixels which are located in the vicinity of image. In [5] author describes choosing randomly $n$ pair of image point ($a_i$, $b_i$) and increases the $a_i$ by one, while decreases the $b_i$ by one. The detection was performed by comparing the sum of the difference of $a_i$ and $b_i$ of the $n$ pairs of the points with $2n$. In [6] wetermark is embedded into LSB of color image using the biological color model. But it uses the combination of both frequency and spatial domain. Many of these techniques suffer from the image processing operations like filtering, cropping, sharpening etc. In addition to these, in many methods supports less embedding capacity.

To resolve these problems, we a propose a novel technique that embeds the monochrome image into color cover image using LSB substitution method. The proposed embedding method uses the intensity value of the pixel to embed the watermark. The intensity $I$ of pixel is calculated as average of Red, Green and Blue intensity channels. The watermark bits are substituted into the channels which have the value greater than intensity $I$ using LSB substitution method. As the color pixel is decomposed into three channels we can achieve the embedding capacity of 3 bits/pixel. But in most of the color images use of normal LSB with payload of 3 bits/pixel produces perceptual poor quality of images and suffers from filtration. To resolve these problems we suggest a method that selects the channels for embedding watermark based on the intensity value of current pixel. Using this method we can achieve the best embedding capacity of 3 bits/pixel and worst embedding capacity of 2bits/pixel. The security of watermark is achieved by permuting the watermark bits using secret key. Before embedding the watermark, the watermark bits are permuted using secret key array. Then in extraction same key array is used to extract the watermark.

————————————————

- *Nagaraj V. Dharwadkar is the PhD research Scholar, in Computer Science and Engineering Department, National Institute of Technology, Warangal, Andhra Pradesh, India.*
- *B. B. Amberker is the Professor in the Department of Computer Science and Engineering, National Institute of Technology, Warangal, Andhra Pradesh, India.*

The remaining paper is organized as follows: In Section 2, we describe the Proposed Algorithm. In Section 3, we describe the experimental results to explain the performance of algorithm. In Section 4 we compare proposed scheme with reference and in Section 5 we conclude our paper.

## 2 PROPOSED ALGORITHM

Normally the color image is represented by the standard 24 bits/pixel or 3 Bytes/pixel. Thus each pixel is represented by 8 bits of each Red, Blue and Green gray scale intensity values. If we use LSB from each channel we can embed the watermark at the most 3 bits/pixel. In the proposed algorithm, to decide the embedding sequence and perceptual similarity between original image and watermarked image, we consider the intensity $I$ of pixel as a parameter. Based on intensity of pixels the channels of pixels are selected dynamically for embedding the watermark. Thus the algorithm embeds the watermark of size is equivalent to color image. The extraction algorithm uses the intensity value of the pixel as a parameter to select the channels for extraction and the watermark bits are extracted from the LSB of the channels. The algorithm is explained in the following sections

### 2.1 Watermark Embedding

Embedding the watermark requires the following steps, and results in an invisible watermark with no visual artifacts.

Algorithm : Watermark Embedding
Input : Color (Cover) Image ($C$) and Monochrome Watermark image ($W$).

1. Read the watermark image $W$ of size $m \times m$,
$$W = \{(i,j) \in \{0,1\} : 1 \leq i \leq m, 1 \leq j \leq m\}$$
2. Preprocess the watermark: Read a binary sequence $K$ which is the secret key.
$$K = \{(i,j) \in \{0,1\} : 1 \leq i \leq m, 1 \leq j \leq m\}$$
and compute $W' = W \oplus K$, where $\oplus$ is bit wise XOR operator
3. Read the color image $C$ of size $m \times m$ and transfer this image into Red ($R$), Green ($G$) and Blue ($B$) channels of size $m \times m$. Then transfer these $R$, $G$, $B$ channels into biological color model Hue ($H$) Saturation ($S$) and Intensity ($I$) using the following equations.

$$H = \begin{cases} \theta & \text{if } B \leq G \\ 360 - \theta & \text{if } B > G \end{cases} \quad (1)$$

$$\theta = \cos^{-1}\left\{ \frac{\frac{(R-G)+(R-B)}{2}}{\left[(R-G)^2 + (R-B)(G-B)\right]^{1/2}} \right\}$$

$$S = 1 - \frac{3}{(R+G+B)} \min(R,G,B) \quad (2)$$

$$I = \frac{(R+G+B)}{3} \quad (3)$$

4. For each pixel select a channel from $R$, $G$, $B$ which is having value $\geq I$ and embed the watermark bit of $W'$ into LSB of selected channels using LSB substitution.
5. Transfer the modified $R$, $G$, $B$ channels into watermarked color image $C'$.

### 2.2 Watermark Extraction

Algorithm : Watermark Extraction
Input : Watermarked Color Image ($C'$) and the secret key ($K$)

1. Read the watermarked color image $C'$ of size $m \times m$ and transfer this image into Red ($R'$), Green ($G'$) and Blue ($B'$) channels of size $m \times m$. Then transfer these $R'$, $G'$, $B'$ channels into biological color model Hue ($H'$), Saturation ($S'$) and Intensity ($I'$) using the equations (1), (2) and (3).
2. Read a binary sequence $K$ which is the secret key.
$$K = \{(i,j) \in \{0,1\} : 1 \leq i \leq m, 1 \leq j \leq m\}$$
3. For each pixel select a channel from $R'$, $G'$, $B'$ which is having value $\geq I'$ and extract the watermark bit $W'$ from LSB of selected channels.
4. Process the $W'$ with key array $K$ to get the extracted watermark $W$ as follows: $W = W' \oplus K$, where $\oplus$ is bit-wise XOR operator.

## 3 EXPERIMENTAL RESULTS

To verify the effectiveness of the proposed method, a series of experiments were conducted. In these experiments, a set of original images of size $256 \times 256$, with 8 bit color representation is used. Fig. 1 shows color images of Lena, House, Girl, Ledy and Peppers which are used in our eperiments. The watermark used is a monochrome image of size $256 \times 256$.

Fig. 2 shows the watermarked color images obtained by our algorithm. To measure the distortion incorporated by the watermarking algorithm we have used Mean Square Error ($MSE$) and Peak Signal to Noise Ratio ($PSNR$). For color image with color components $R$, $G$, and $B$ the $MSE$ and $PSNR$ is given by:

$$MSE = \frac{\sum_{K \in \{R,G,B\}} \sum_{i=1}^{M} \sum_{j=1}^{N} \left[ I(i,j)_K - I'(i,j)_K \right]^2}{3MN}$$

Here, $M$ and $N$ are the height and width of image respectively. $I(i,j)$ and $I'(i,j)$ are the $(i,j)^{th}$ pixel value of the original image and modified image respectively.

$$PSNR = \log \frac{(2^n - 1)^2}{MSE}$$

Table 1 shows the *PSNR* and *MSE* between original color image and watermarked images. We observed that the average *PSNR* and *MSE* for the set of images is nearly 60.82 dB and 0.053 respectively. The similarity of extracted and original watermark has been quantitatively measured by the Normalized correlation (*NC*) and Standard correlation (*SC*) which are defined as follows:

$$NC = \frac{\sum_{i=1}^{M} \sum_{j=1}^{N} [I(i,j) \, I'(i,j)]}{\sum_{i=1}^{M} \sum_{j=1}^{N} [I(i,j)]^2}$$

Where *I* (*i*, *j*) is original image and *I'*(*i*, *j*) is modified image, *M* is height of image and *N* is width of image.

$$SC = \frac{\sum_{i=1}^{M} \sum_{j=1}^{N} [I(i,j) - I'][J(i,j) - J']}{\sqrt{\sum_{i=1}^{M} \sum_{j=1}^{N} [I(i,j) - I']} \sqrt{\sum_{i=1}^{M} \sum_{j=1}^{N} [J(i,j) - J']}}$$

Here, *I* (i, j) is original watermark, *J* (i, j) is extracted watermark, *I* ' is the mean of original watermark and *J* ' is mean of extracted watermark.

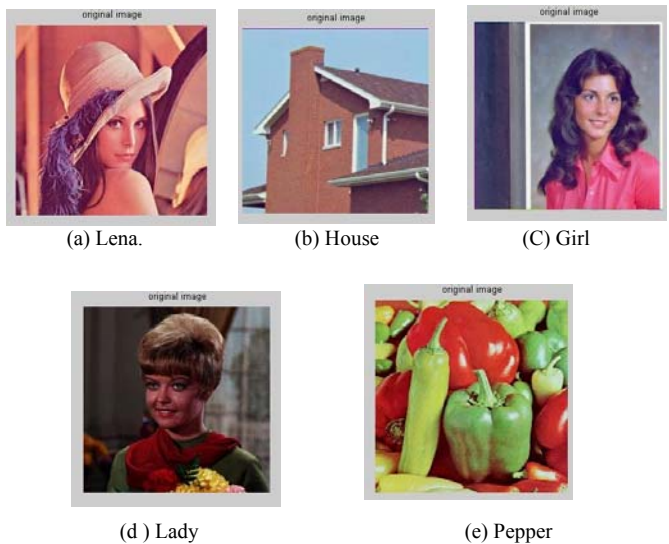

Fig. 1. Original color images.

Fig. 3 shows the original watermark, permuted watermark and extracted set of watermarks. The Table 2 shows the NC and *SC* value between original and extracted watermark for set of color images. For all images the *NC* and *SC* values between original and extracted watermark are nearly equal to 1, which shows that extracted watermark is tightly correlated to original watermark.

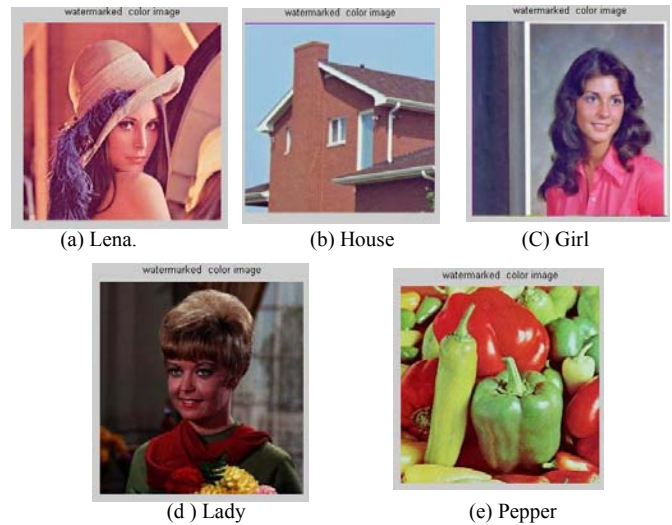

Fig. 2. Watermarked color images

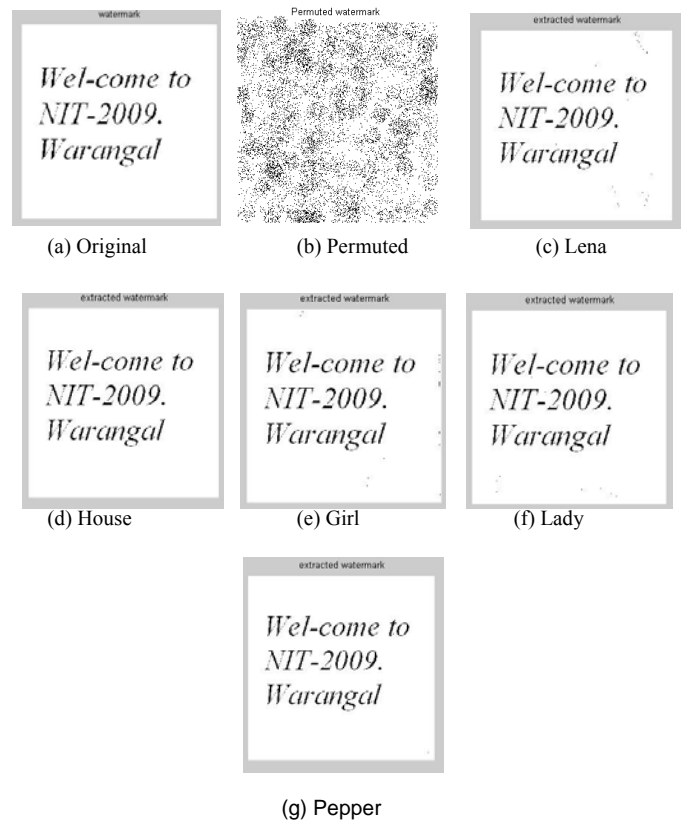

Fig. 3. Original watermark, permuted watermark and extracted watermarks from different Watermarked color images

TABLE 1
MSE AND PSNR BETWEEN ORIGINAL COLOR IMAGE AND WATERMARKED COLOR IMAGE

| Properties | Lena | House | Girl | Lady | Pepper |
|---|---|---|---|---|---|
| MSE | 0.053 | 0.053 | 0.053 | 0.054 | 0.053 |
| PSNR | 60.86 | 60.82 | 60.82 | 60.80 | 60.84 |

TABLE 2
NC AND SC BETWEEN EXTRACTED WATERMARK AND ORIGINAL WATERMARK

| Properties | Lena | House | Girl | Lady | Pepper |
|---|---|---|---|---|---|
| NC | 0.99 | 1.0 | 0.99 | 0.99 | 1.0 |
| SC | 0.99 | 1.0 | 0.99 | 0.99 | 0.99 |

### 3.1 Effect of attacks

The performance of proposed algorithm is analyzed by considering image processing attacks like cropping, filtration and compression.

**1. Cropping:** The watermarked image is cropped in terms of percentage of image size. The effect of cropping is decided calculating the NC and SC between extracted watermark and original watermark. Fig. 4 shows the 20% cropped watermarked Pepper image and extracted watermark. Thus from Fig 5 it is cleared that our proposed watermarking algorithm is robust for cropping on the set of watermarked images. The algorithm is rigid for upto 40 % cropping.

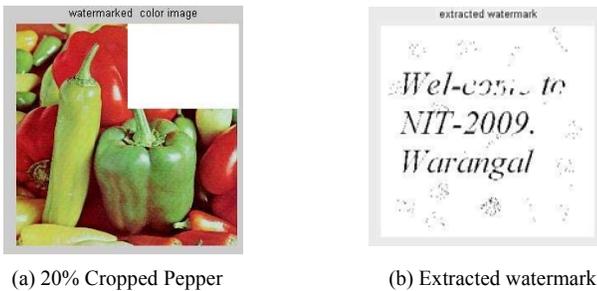

(a) 20% Cropped Pepper     (b) Extracted watermark

Fig. 4. Effect of Cropping

**2. Compression:** The watermarked image is compressed with different Quality factors. The effect of compression is observed by calculating the NC and SC between extracted watermark and original watermark. Fig. 6 shows the effect of compression on set of watermarked images. As the compression factor increases the quality of extracted watermark decreases. Thus our algorithm is robust against compression of quality factor upto 50.

**3. Image Blurring:** Special type of circular averaging filter is applied on the watermarked color image to analyze the effect of Blurring. The circular averaging (pillbox) filter filters the watermarked image within the square matrix of side *2× (Disk Radius) +1*. The disk radius is varied from 0.5 to 1.4 and the effect of blurring is analyzed on extraction algorithm.

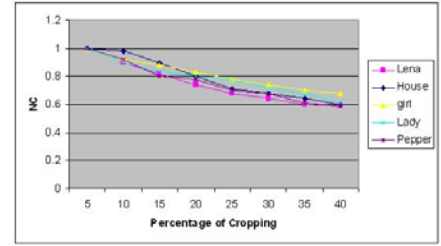

(a) NC

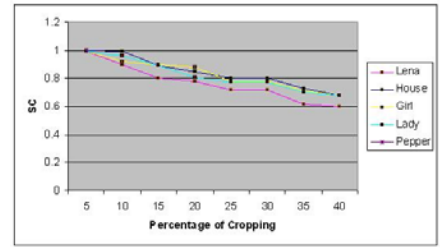

(b) SC

Fig. 5. Effect of cropping measured by NC and SC between the original and the extracted watermark on the set of sample images

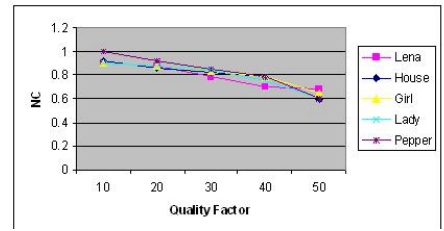

(a)    NC

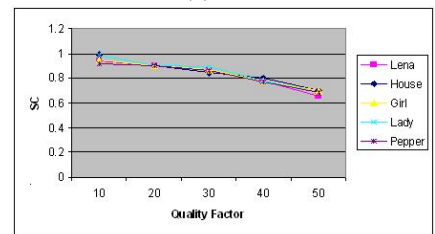

(b) SC

Fig. 6. Effect of compression measured by NC and SC between the original and the extracted watermark on the set of sample images.

Performance of the algorithm is measured by calculating *NC* and *SC* between the extracted watermark and the original watermark. Fig 7. Shows the effect of blurring on the set of watermarked images. As the disk radius of the filter is increased and applied on the watermarked image, the quality of extracted watermark decays.

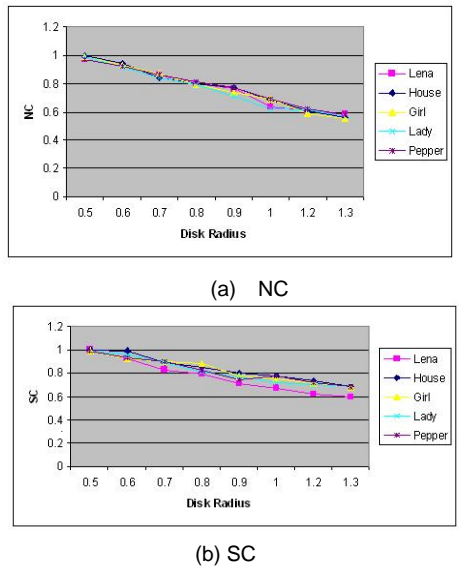

(a) NC

(b) SC

Fig. 7. Effect of Blurring measured by NC and SC between the original and the extracted watermark on the set of sample images

**4. Salt and Pepper noise:** The salt and pepper noise is added to the watermarked image $I$. This affects approximately $d \times (size(I))$ pixels, where $d$ is the noise density. The performance of extraction algorithm is analyzed by increasing density of the noise starting from 0.001 up to 0.007. The extracted watermark and original watermark are compared in terms of $NC$ and $SC$. Fig. 8 shows the effect of salt and pepper noise on the set of watermarked images. The increase in the noise density reduces the performance of extraction algorithm. Thus the correlation between the extracted and original watermark decreases as the density of noise increases.

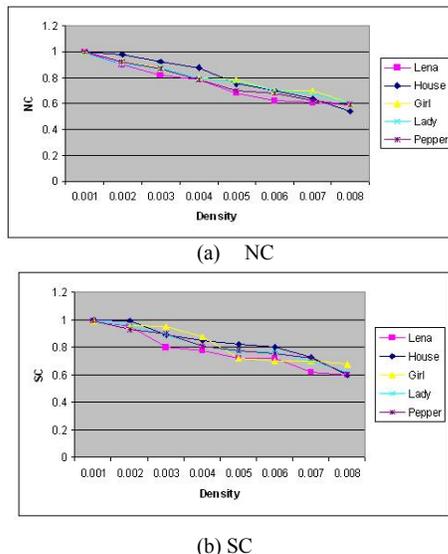

(a) NC

(b) SC

Fig. 8 Effect of salt and Pepper Noise measured by NC and SC between the original and the extracted watermark on the set of sample images

## 4 COMPARISON

We compare the performance of our watermarking algorithm with the Nino. D[6] which also uses the Biological color model. The comparison is shown in Table 3. In the proposed algorithm the watermark is embeded directly into the LSB of channels of the pixel based on intensity of pixel, without using additional transformations. However, the method proposed in [6] uses dual domain for embedding watermark.

TABLE 3
COMPARISON OF PROPOSED ALGORITHM WITH NINO. D [6]

| Properties | Nino. D [6] | Proposed Algorithm |
|---|---|---|
| Processing Domain | Spatial and Frequency | Spatial Domain |
| Cover Data | Color Image | Color Image |
| Embedding type | Block Wise | Pixel Wise |
| Embedding Capacity | Less than the size of cover image | Greater than the size of cover image |
| Watermark | Binary Image | Monochrome Image |
| security of watermark | Hadmard Transformation | Secret Key (Array) |
| PSNR | ------------------------- | 60.28dB |
| Color model considered | RGB and HSI | RGB and HSI |
| Effect of Attacks Analyzed | Compression, Gaussian noise, Rotation, salt and pepper Noise | Compression, Cropping, Sharpening, salt and pepper Noise |

## 5 CONCLUSION

The proposed algorithm is blind invisible spatial based watermarking algorithm. This method embeds the watermark into the pixels of image by considering the intensity of pixels. Security to the watermark is provided using secret key. Security of watermark can be further enhanced by using hash functions or chaotic binary sequence. The effect of cropping and filtration on watermark can be minimized by considering the texture properties of pixels. The proposed algorithm is weak to geometrical transformation. Thus the futer work is to design a geometrical transform-invariant watermarking algorithm.

**Nagaraj V. Dharwadkar:** He has obtained B.E. in Computer Science & Engineering in 2000 and M.Tech.in Computer Science & Engineering in 2006. He is PhD Scholar in Computer Science and Engineering at National Institute of Technology, Warangal, AP, India. His research area of interest is Digital Watermarking and Visual Cryptography

**B. B. Amberker:** He obtained his Ph.D from the Department of Computer Science and Automation, IISc. Bangalore, India. He is presently working as Professor in the Department of Computer Science and Engineering, National Institute of Technology, Warangal, AP, India. His research areas of interest are Computational Number Theory and Cryptography, Authentication Schemes and Secure Group Communication.